\input harvmac
\noblackbox
\input epsf.tex
\overfullrule=0pt
\def\Title#1#2{\rightline{#1}\ifx\answ\bigans\nopagenumbers\pageno0\vskip1in
\else\pageno1\vskip.8in\fi \centerline{\titlefont #2}\vskip .5in}

\font\ticp=cmcsc10

\font\secfont=cmcsc10

\font\mytitlefont=cmbx12

\font\cmss=cmss10 
\font\cmsss=cmss10 at 7pt
\newcount\figno
\figno=0
\def\fig#1#2#3{
\par\begingroup\parindent=0pt\leftskip=1cm\rightskip=1cm\parindent=0pt
\baselineskip=11pt
\global\advance\figno by 1
\midinsert
\epsfxsize=#3
\centerline{\epsfbox{#2}}
\vskip 12pt
{\bf Fig.\ \the\figno: } #1\par
\endinsert\endgroup\par
}
\def\figlabel#1{\xdef#1{\the\figno}}
\def\encadremath#1{\vbox{\hrule\hbox{\vrule\kern8pt\vbox{\kern8pt
\hbox{$\displaystyle #1$}\kern8pt}
\kern8pt\vrule}\hrule}}

%
%
\baselineskip=18pt plus 2pt minus 2pt

%
\def\CH{{\cal H}}

\def\IZ{\relax\ifmmode\mathchoice
{\hbox{\cmss Z\kern-.4em Z}}{\hbox{\cmss Z\kern-.4em Z}}
{\lower.9pt\hbox{\cmsss Z\kern-.4em Z}}
{\lower1.2pt\hbox{\cmsss Z\kern-.4em Z}}\else{\cmss Z\kern-.4em }\fi}
\def\IC{\relax\hbox{$\inbar\kern-.3em{\rm C}$}}
\def\IR{\relax{\rm I\kern-.18em R}}
\def\bZ{\bf Z}

\def\zb{\bar z}
\def\zt{{\tilde z}}
\def\yt{{\tilde y}}
\def\xt{{\tilde x}}
\def\ft{{\tilde f}}
\def\gt{{\tilde g}}
%

\def\z{ z}
%


\def\r{\rho}

\def\s{\sigma}

\def\t{\tau}
\def\a{\alpha}

\def\p{\pi}

\def\O{{\Omega}}
\def\D{\Delta}

%


\def\1{\relax 1 { \rm \kern-.35em I}}

%

\def\frac#1#2{{#1 \over #2}}

\def\p+{{\partial_+}}

\def\half{{1 \over 2}}
\def\ket#1{|#1\rangle}


\def\apm{\alpha^{\prime}}

\def\zb{\bar{z}}
\def\[{\left [}
\def\]{\right ]}
\def\({\left (}
\def\){\right )}
\def\aD{{\bar{D}7}}
\def\aO{{\bar{O}7}}
\def\aDf{{\bar{D}5}}
\def\aOf{{\bar{O}5}}
\def\FL{(-1)^{F_L}}
\def\FR{(-1)^{F_R}}

\def\OL{\Omega (-1)^{F_L}}
\def\OR{\Omega (-1)^{F_R}}
\def\OF{\Omega (-1)^{F}}
\def\II{ I_{89}}
\def\hs{\sigma}
\def\TrH#1{ {\raise -.5em
                      \hbox{$\buildrel {\textstyle  {\rm Tr } }\over
{\scriptscriptstyle \CH _ {#1}}$}~}}

\Title{\vbox{\baselineskip12pt
\hbox{\ticp TIFR/TH/01-35}
\hbox{hep-th/0109019}
}}
{\vbox{\centerline {\mytitlefont ON CONDENSATION OF CLOSED-STRING TACHYONS} }}
\centerline{{\ticp
Atish Dabholkar
}}

\vskip.1in
\centerline{\it Department of Theoretical Physics}
\centerline{\it Tata Institute of Fundamental  Research}
\centerline{\it Homi Bhabha Road, Mumbai, India 400005.}
\centerline{Email: atish@tifr.res.in}
\vskip .1in

\bigskip
\centerline{ABSTRACT}
\medskip

An F-theory dual of a nonsupersymmetric orientifold is considered. It
is argued that the condensation of both open and closed string
tachyons in the orientifold corresponds to the annihilation of branes
and anti-branes in the F-theory dual. The end-point of tachyon
condensation is thus expected to be the vacuum of Type-IIB
superstring. Some speculations are presented about the F-theory dual
of the bosonic string and tachyon condensation thereof.

\bigskip

\bigskip
\Date{September 2001}

\vfill\eject

\def\ajou#1&#2(#3){\ \sl#1\bf#2\rm(19#3)}
\def\npb#1#2#3{{\sl Nucl. Phys.} {\bf B#1} (#2) #3}
\def\plb#1#2#3{{\sl Phys. Lett.} {\bf B#1} (#2) #3}

\def\prd#1#2#3{{\sl Phys. Rev. }{\bf D#1} (#2) #3}

\def\prep#1#2#3{{\sl Phys. Rep. }{\bf #1} (#2) #3}

\def\jhep#1#2#3{{\sl JHEP}{\bf #1} (#2) #3}

%

\lref\Sen{A. Sen, {\it D-branes as Solitons},
 http://online.itp.ucsb.edu/online/mp01/sen1/.}

\lref\SenI{A. Sen, 
{\it Tachyon Condensation on the Brane Anti-brane System},
\jhep{9808}{1998}{12}, hep-th/9805170.}

\lref\SenII{A. Sen, 
{\it F-theory and Orientifolds}, \npb{475}{1996}{562},
hep-th/9605150.}

\lref\SeWi{N. Seiberg, E. Witten, 
{\it Monopoles, Duality and Chiral Symmetry Breaking in $N=2$
Supersymmetric QCD}, \npb{431}{1994}{484}, hep-th/9408099}

\lref\BDS{T. Banks, M. R. Douglas, N. Seiberg, 
{\it Probing F-theory with Branes},\plb{387}{1996}{278},
hep-th/9605199.}

\lref\BaSu{T. Banks and L. Susskind, {\it Brane-Antibrane Forces}, 
hep-th/9511194.}

\lref\Vafa{C. Vafa, {\it Evidence for F-theory}, 
\npb{469}{1996}{403}, hep-th/9602022.}

\lref\MoVa{D. Morrison and C. Vafa, 
{\it Compactifications of F-theory on Calabi-Yau Three-folds I, II},
\npb{B473}{1996}{74}, hep-th/9602114;
\npb{476}{1996}{437}, hep-th/9603161.}

\lref\GSVY{B. Greene, A. Shapere, C. Vafa, and S.~T.~Yau, 
{\it Stringy Cosmic Strings and Non-compact Calabi-Yau Manifolds},
\npb{337}{1990}{1}.}

\lref\APS{A. Adams, J. Polchinski, E. Silverstein, 
{\it Don't Panic! Closed String Tachyons in ALE Spacetimes},
hep-th/0108075.}

\lref\ADS{I. Antoniadis, E. Dudas, and A. Sagnotti, 
{\it Supersymmetry-breaking, Open Strings and M-theory}, 
\npb{474}{1996}{361}, hep-th/9807011.}

\lref\KKS{S. Kachru, J. Kumar, and E. Silverstein, 
{\it Orientifolds, RG Flows, and Closed-string Tachyons}, 
{\sl Class. Quant. Grav.} {bf 17} (2000) 1139, hep-th/9907038.}

\lref\BeGa{O. Bergman and M. Gaberdiel, 
{\it A Non-supersymmetric Open String Theory and S-Duality},
\npb{499}{1997}{183}, hep-th/9701137.}

\lref\BeGaI{O. Bergman and M. Gaberdiel, 
{\it Dualities of Type-0 Strings}, \jhep{9907}{1999}{022},
hep-th/9906055.}

\lref\CoGu{M. S. Costa, M.Gutperle, 
{\it The Kaluza-Klein Melvin Solution in M-theory}, 
\jhep{0103}{2001}{27}, hep-th/0012072.}

\lref\GuSt{M. Gutperle, A.Strominger, 
{\it Fluxbranes in String Theory}, \jhep{0106}{2001}{035},
hep-th/0104136.}

\lref\Suya{T. Suyama, 
{\it Closed String Tachyons in Non-supersymmetric Heterotic Theories},
hep-th/0106079.}

\lref\DiHa{L.~J.~Dixon, J.~A.~Harvey, 
{\it String Theories in Ten Dimensions Without Spacetime
Supersymmetry}, \npb{274}{1986}{93}.}
 
\lref\AGMV{L. Alvarez-Gaume, P. Ginsparg, G. Moore, C. Vafa, 
{\it An $O(16)\times O(16)$ Heterotic String}, \plb{171}{1986}{155}.}

\lref\SeWiI{N. Seiberg, E. Witten, 
{\it Spin Structures in String Theory}, {\sl Nucl.\ Phys. } {\bf B276}
(1986) 272.}

\lref\KLT{H. Kawai, D. C. Lewellen, S. H. Tye, 
{\it Classification of Closed-fermionic-string Models},
\prd{34}{1986}{3794}.}

\lref\BlDi{J. D. Blum, K. R. Dienes, 
{\it Duality without Supersymmetry: The Case of the SO(16) $\times$
SO(16) String}, \plb{414}{1997}{260}, hep-th/9707148; \ {\it
Strong/Weak Coupling Duality Relations for Non-Supersymmetric String
Theories}, \npb{516}{1998}{83}, hep-th/9707160.}

\lref\Dabh{A. Dabholkar, {\it Lectures on Orientifolds and Duality}, 
{\sl Proceedings of the Summer School of High Energy Physics, ICTP,
Trieste, Italy, 1997}, hep-th/9804208.}

\lref\CENT{A. Casher, F. Englert, H. Nicolai, and A. Taormina,
{\it Consistent Superstrings as Solutions of the $D=26$ Bosonic String 
Theory}, \plb{162}{1985}{121}.}

\lref\LSW{W. Lerche, A. Schellekens, and N.~P.~Warner, 
{\it Lattices and Strings}, \prep{177}{1989}{1}.}

\lref\DaIq{{\it work in progress}}

\newsec{\secfont Introduction and Summary}

Much has been learnt recently about the condensation of open-string
tachyons (see \Sen\ for an overview). One of the main ingredients that
has made it possible to successfully analyze this very off-shell
process is the fact that even before doing any computation there is a
compelling {\it a priori} argument about the end-point of tachyon
condensation. The open-string tachyon in a brane-antibrane system has
an interpretation in the closed-string channel as the instability
towards brane-antibrane annihilation \BaSu.  Therefore, the end-point
of tachyon condensation is expected to be nothing but the closed
string vacuum.  This physical picture underlies the `Sen conjecture'
about the height of the tachyon potential \SenI\ and has been a very
valuable guide for the explicit computations of the tachyon potential
using various off-shell formalisms.

One of the difficulties in applying similar ideas to the more
interesting case of the closed-string tachyon is that there is no
analogous dynamical picture for the condensation of closed-string
tachyons that is immediately obvious.

In this paper, we address this issue for a nonsupersymmetric
orientifold model in eight dimensions. Before presenting the details,
let us summarize the main argument. The orientifold theory that we
consider contains orientifold 7-planes as well as anti-orientifold
7-planes. Tadpoles are canceled by including both $D7$-branes and
$\aD$-branes. The spectrum contains tachyons in certain regions of the
moduli space from both the open-string and the closed-string sectors.
The open-string tachyons indicate as above the instability towards the
annihilation of $D7$-$\aD$ branes and arise when the separation
between branes and anti-branes is sufficiently small. However, there
is no analogous interpretation of the closed string
tachyons. Moreover, the $O7$ and the $\aO$ planes both have negative
tension and hence are not dynamical objects. In the perturbative
orientifold theory they cannot get annihilated and will stay around
even after the annihilation of $D7$-$\aD$ branes. Nonperturbatively,
it is natural to consider an F-theory dual of this orientifold theory
in which each $O7$-plane is replaced by a pair of $(p, q)$
7-branes. The $\aO$ plane is similarly replaced by a charge-conjugate
pair of $(p, q)$ 7-branes. Now, in the F-theory picture a $(p, q)$
7-brane is a dynamical object locally not any different form a
$D7$-brane and can certainly get annihilated along with its charge
conjugate $(-p, -q)$ 7-brane. Thus, a collection of $(p, q)$ branes
and antibranes is  unstable towards annihilation. We
therefore interpret the various open and closed string tachyons of the
orientifold as a signal of this instability. In these models, there
are equal number of $(p, q)$ 7-branes and anti 7-branes so they can
annihilate each other completely emitting various closed string modes
such as gravitons and dilatons.  This process will leave behind the
vacuum of the Type-IIB superstring.  We are thus led to the conjecture
that the condensation of open and closed tachyons of the orientifold
leads to the supersymmetric Type-IIB string as the endpoint.

We present the details of the orientifold in \S2
and discuss the proposed F-theory dual in \S3. We conclude in \S4
with some speculative remarks about the closed-string tachyons in the
bosonic string.

\newsec{{ \secfont A Non-supersymmetric Orientifold}}

Our starting point will be an orientifold of the Type-IIB string
compactified on a 2-torus to eight dimensions by a $\bZ_2\times \bZ_2$
symmetry $\{ 1, \OL\II\}\times\{ 1, \OR\II\hs\}$ where $\O$ is the
usual orientation reversal on the worldsheet, $\II$ is the reflection
of the $89$ coordinates, $\hs$ is a half-shift along the 9th
direction, and $F_L$ and $F_R$ are the left and right-moving spacetime
fermion numbers respectively \refs{\ADS,\KKS}.

Let us first look at the spectrum in the closed-string sector. The
massless states from the untwisted sector that survive the orientifold
projections are the metric, the dilaton and the 2-form RR field. The
orientifold group $\{ 1, \OL\II, \OR\II\hs, (-1)^F\hs\}$ contains the
element $(-1)^F\hs$ where $F=F_L +F_R$ is the total fermion number;
hence the spectrum will contain states twisted by this element. In the
twisted sector, the ground state energy in the light-cone
Green-Schwarz formalism is $-{8\over 24} -{8\over 48}=-\half$ for both
left-movers and right-movers because eight fermions are half-integer
moded and eight bosons are integer moded. If $(-1)^F$ were not
accompanied by the half-shift $\hs$ around the circle along the 9th
direction, this would lead to a neutral closed-string tachyon as well
as an additional massless R-R 2-form. However, because of the
half-shift, all these states have large positive mass-squared when the
radius of the circle along the 9th direction is large. Only when this
radius becomes smaller than the string-length does one see a tachyon
in the spectrum.  All closed-string massless fermions are projected
out.

In the open-string sector, there are four orientifold 7-planes at the
fixed points of $ \OL\II$. Moreover, there are four anti-orientifold
7-planes at the fixed points of $\OR\II\hs$ For simplicity, let us
consider a square torus with the radii of the circles in the 8 and 9
directions to be $R$. Let us write $z=X^8 + i X^9$ as the complex
coordinate of the torus. Then the orientifold planes are located at
the fixed points of $\II$:$z_1=0$: $z_2=\half$, $z'_1=0+{i\over2}$,
and $z'_2=\half + {i\over 2}$ in units of $R$.  Similarly, the
anti-orientifold planes are located at $z_3=0+{i\over 4}$, $z_4=\half+
{i\over 4}$, $z'_3=0+{3i\over 4}$ and $z'_4=\half+ {3i\over 4}$ which
are the fixed points of $\II\hs$. Note
that $z'_1$ and $z'_2$ respectively are the images of $z_1$ and $z_2$
under $\II\hs$ whereas $z'_3$ and $z'_4$ respectively are the images
of $z_3$ and $z_4$ under $\II$. (Fig 1)\fig{The crosses indicate the
orientifold planes and the circles indicate anti-orientifold planes.
The fundamental region of the orientifold is the quarter of the
original torus that is inside dashed lines.}{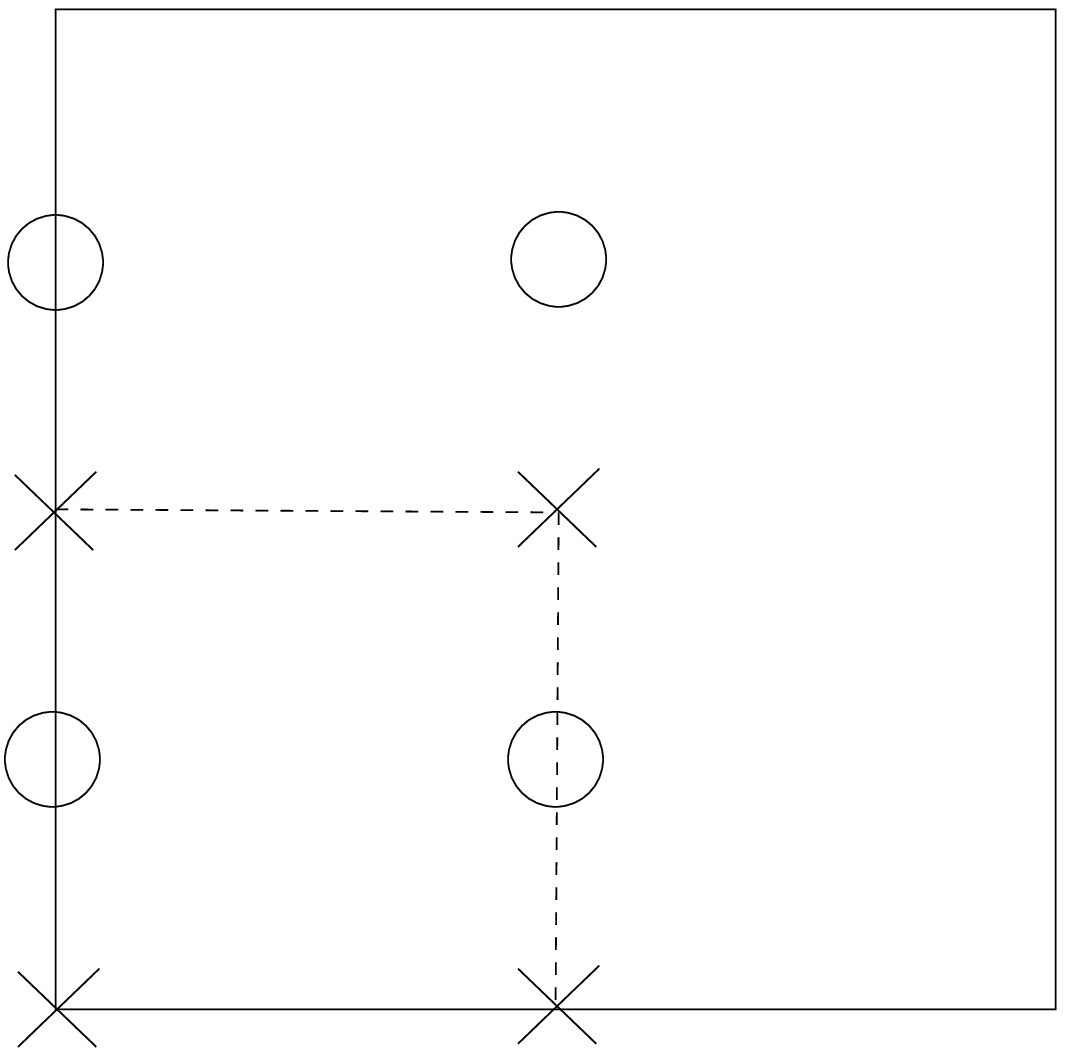} {1.0truein}
\noindent The tadpoles can be canceled locally by putting eight
elementary D7-branes at each of $O7$-planes and eight $\aD$-branes at
each of the $\aO$ planes\footnote{$^{\dag}$}{By elementary D-branes we mean
the D-branes of the original Type-IIB string before the orientifold
projection. On the $\bZ_2\times \bZ_2$ orientifold we need four
elementary branes to make a single dynamical D-brane that can move
away from the orientifold planes.}. The spectrum contains $SO(8)^4$
gauge bosons and massless fermions in the adjoint reps of the gauge
group. There are charged scalar multiplets that coming from the
$7-{\bar 7}$ open strings that become tachyonic when the separation
between the branes and anti-branes is sufficiently small.
    
At tree-level, there are a number of massless moduli fields.  We can
talk about the classical moduli space even though these moduli will
typically be lifted at loop level and moreover in the interacting
theory the tachyons will in any case destabilize the vacuum. In the
closed-string sector, there are two complex moduli that correspond to
complex-structure and K\"ahler deformations and a real modulus that
corresponds to the dilaton. The open-string sector contributes sixteen
complex moduli corresponding to the positions of the sixteen dynamical
D-branes on the torus. There are thus altogether eighteen complex
moduli and one real modulus.

If we move the D-branes away from the orientifold planes then the
gauge symmetry is broken to appropriate subgroups of
$SO(8)^4$. However, in this case the tadpoles are not canceled locally
even though the total charge on the torus is zero. As a result, the
metric of the torus as well as the dilaton would vary as we move around
on the torus.

Before discussing the F-theory dual, we would like to note that there
are two other $\bZ_2\times \bZ_2$ orientifolds that have been
considered in the literature in various contexts.  The orientifold
action is closely related and but the spectrum is distinct.

\item{A.} The orientifold group is
$\{ 1, \OL\II\}\times\{ 1, \OR\II\}$. Here there is no half-shift in
the orientifold action and as a result the there are four pairs of
coincident $O7$-$\aO$ planes. The gauge symmetry $SO(8)^8$ is twice as
large and there are tachyons at all points in the moduli-space. This
is T-dual to the orientifold $\{ 1, \O\}\times\{ 1, \OF\}$ which has been
conjectured in \BeGa\ to be S-dual to the 26-dimensional bosonic
string compactified to ten dimensions on the $Spin (32)$ lattice. We will
comment upon this orientifold in \S4.

\item{B.} The orientifold group is 
$\{ 1, \OL\II\}\times\{ 1, \OR\II{\tilde\hs}\}$. The half-shift
${\tilde\hs}$ is T-dual to $\hs$\footnote{$^{\dag}$}{On a state $\ket{m,
n}$ with quantized momentum $m$ and winding $n$ along the 9th
direction, $\s$ acts with a phase $(-1)^m$ whereas ${\tilde\hs}$ acts
with the phase $(-1)^n$.}. The spinors no longer transform in the
adjoints but rather in the bi-vector representations of
$SO(8)^4$. Under T-duality along both the 8 and 9 directions $\OL\II$
goes to $\O$ and $\tilde\hs$ goes to $\hs$. The T-dual orientifold
symmetry is equivalently $\{1, \O\}\times \{1, \O (-1)^F\hs\}$. This
is an orientifold of a variant of the 0B string considered in \BeGaI\ and is
conjectured in \BlDi\ to be dual to the $SO(16)\times SO(16)$
heterotic string \refs{\DiHa, \AGMV, \KLT}.

\newsec{\secfont Nonsupersymmetric F-theory Dual}

The F-theory dual of our orientifold can be established reliably
at the special points the moduli space with $SO(8)^4$ symmetry where
the tadpoles are canceled locally. Our reasoning will be analogous to
that used for establishing the supersymmetric duality between Type-I
and F-theory \SenII. However, at a generic point in the moduli space
as the branes move around, the F-theory dual will
be valid only in the approximation that the branes and anti-branes are
far apart from each other such that their interaction with each other
can be ignored.  The approximation and the regime of its validity will
become clear below.

Our starting point will be an elliptic curve in the Weierstrass form
\eqn\fone{
y^2 = x^3 + x f(z, \zb) + g(z, \zb)}
where $x,y,z \in \bf{CP^1}$, $f(z, \zb)$ is a polynomial of combined
degree eight and $g(z, \zb)$ is a function of combined degree twelve
in $z, \zb$. Unlike in the supersymmetric case \refs{\Vafa,
\MoVa}, the curve depends on both the holomorphic and anti-holomorphic 
coordinates $(z, \zb)$ of the base. An explicit parametrization of
these functions suitable for our purposes is given later. This
equation defines an elliptic fibration over an $\bf S^2$ base
coordinatized by $(z, \zb)$. The $\tau$ parameter of the elliptic
fiber is determined by
\eqn\modular{
j(\tau(z,\zb))= {{4. (24f)^3}\over{27g^2+4f^3}}
}
where $j(\tau)$ is the usual modular invariant function normalized so
that $j(i)=(24)^3$. For large values of imaginary part of $\tau$ we
have,
\eqn\jfunction{
j(\tau) \sim e^{-2\pi i\tau}.}

Now, we define the nonsupersymmetric F-theory as the  compactification of 
Type-IIB theory on the non-Ricci-flat $\bf S^2$ base coordinatized by
$(z, \zb)$. The axion-dilaton field as a function of the coordinates
of the base is given by the modular parameter of the above curve:
\eqn\axion{
a(z, \zb) + i e^{-\phi(z, \zb)} \equiv \tau(z, \zb),}
where $a$ is the R-R scalar and $\phi$ is the dilaton.

The fiber degenerates generically at 24 points where the discriminant
\eqn\discri{
\D \equiv 4f^3+27g^2}
vanishes. There are twelve `zeroes' and twelve `anti-zeroes'. Near a
zero $z=z_i$,
\eqn\zero{
j(\tau(z, \zb) \sim {1\over z-z_i}} 
and therefore the axion-dilaton field is given by
\eqn\zerotau{
\tau(z,\zb) \sim {1\over 2\pi i}\ln(z-z_i).}
When $z$ goes around $z_i$, the monodromy of the axion-dilaton field
is given by $\tau \rightarrow \tau +1$, which means that there is a
single 7-brane located at the zero that carries one unit of magnetic
charge with respect to the axion field. The metric has a conical
singularity with deficit angle $\pi/6$ as for a stringy cosmic string
\GSVY. Similarly near an `anti-zero' $\zb =\zb_j$, we have,
\eqn\azero{
j(\tau(z, \zb) \sim {1\over \zb-\zb_j}} 
and
\eqn\azerotau{
\tau(z,\zb) \sim {1\over 2\pi i}\ln(\zb-\zb_j).}
Thus, inclusion of anti-branes on the base necessarily requires that
the axion-dilaton field has the anti-holomorphic dependence on
$\zb$\footnote{$^{\dag}$}{Note that $\tau(z,\zb) \sim -{1\over 2\pi
i}\ln(z-z_i)$ also has the monodromy appropriate for an anti-brane,
however, in this case the imaginary part of $\t$ is no longer positive
definite as required physically from the definition of $\tau$.}. We
expect that to the F-theory defined using the equation \fone would be
a good approximation so long as the parameters are chosen such that
the zeroes and anti-zeroes are well separated.  In this approximation,
the total deficit angle of twelve branes and twelve anti-branes adds
up to $4\pi$, so the base is a sphere with Euler character two. Here,
because we are dealing with a non-supersymmetric situation, we talk
about a real $\bf S^2$ and the Euler character instead of $\bf CP^1$
and first Chern class.

To make contact with the orientifold, we consider the theory at the
special point in the moduli space with $SO(8)^4$ symmetry, where the
tadpoles are canceled locally. In the Type-IIB language, this implies
that the axion-dilaton field, $\tau$, is a constant on the torus and
the metric is flat except for four conical singularities at the
orientifold fixed points with conical deficit angle of $\pi$ at each
singularity.  To obtain the F-theory dual we choose
\eqn\functions{\eqalign{
   f(z,\zb) =&\, \alpha \Phi(z,\zb)^2, \quad g(z, \zb) = \Phi(z,\zb)^3, \cr
  \Phi(z,\zb)=&\, (z-z_1)(z-z_2)(\zb-\zb_3)(\zb-\zb_4)\cr }}
where $z_i$'s are
the locations of the fixed points as defined in the previous section
and $\alpha$ is a constant.   The discriminant now 
\discri\ has two zeroes of order six at the points $z_1, z_2$ implying that 
there are six coincident 7-branes sitting at each of these points. Similarly,
there are two anti-zeroes of order six at the points $\zb_3, \zb_4$
signifying six anti 7-branes at each of these points.  
Now, since $f^3/g^2 = \alpha^3$ is a constant, the $j$-function is a
constant from \jfunction, and therefore the $\tau$ field is a constant
over the base; 
however, there is an $SL(2, \bZ)$ monodromy around each of the points
$z_1, \ldots, \z_4$
\eqn\monodromy{
\left(\matrix{-1&0\cr
               0&-1\cr}\right).}

The metric on the base is given, in this limit, by 
\eqn\metric{
ds^2 = {dzd\zb \over \prod_i|(z-z_i)|}}
up to over all
normalization \GSVY.

The metric is thus flat everywhere except at four points. At each of
these points there is a deficit angle of $\pi$ corresponding to a
bunch of coincident six branes or six anti-branes. Note that the
monodromy around two bunches of six branes is the square of the
monodromy matrix in \monodromy\ which is identity. Thus, there is no
net monodromy and as a result the configuration of twelve branes and
twelve anti-branes can be patched together smoothly. 

It is interesting to note that there is {\it no net force} between a
bunch of twelve branes and a bunch of twelve anti-branes in this
particular non-BPS configuration even though there is an attractive
force between a single brane and a single anti-brane.  This behavior
is much better than what one might naively expect for an assembly of
oppositely charged objects such as electrons and positrons which would
experience an attractive force at any distance. It is possible because
the charges of 7-branes in Type-IIB are non-abelian and as a result
this nonsupersymmetric configuration where tadpoles cancel is locally
stable. In particular, there is no tachyon all the way up to small
values of the radius $R$. The F-theory description in terms of
Type-IIB supergravity is of course valid only when the base of the
compactification is large compared to the string scale.

The local analysis near the singularity is identical to the
supersymmetric case.  For example, near the point, $\zb =\zb_3$, the
curve takes the form of a $D_4$ singularity 
\eqn\near{
\yt^2 \sim \xt^3 +\xt\,\a{\zt}^2 + {\zt^3}}
after the rescaling
\eqn\rescale{
\yt = y\, \phi^{3/2},\quad \xt =x\, \phi,\quad \zt = (z-z_1)}
with 
\eqn\resca{
\phi = (z_1-z_2)(\zb_1-\zb_3)(\zb_1 -\zb_4).}
This $D_4$ singularity corresponds to enhanced $SO(8)$ symmetry as in
the supersymmetric case and we get $SO(8)^4$ symmetry from the four
singular points $z_1, \ldots, z_4$.

Within F-theory, we can also consider more general configurations in
which the D-branes are moved away from the orientifold plane. For
example, locally, the deformation of \near\ away from the $D_4$
singularity is given by the equation
\eqn\deform{
y^2 = x^3 + x\, f({z})  + g({z}) }
where $f$ and $g$ are now polynomials of degree two and three
respectively. To begin with, the equation will depend on the seven
complex parameters of the polynomials of which one can be removed by a
shift of $\zb$ and one more by rescaling $x$ and $y$. The remaining
five parameters are related, as explained in \SenII, to the parameters
of Seiberg-Witten curve for the $N=2$ supersymmetric $SU(2)$ gauge
theory with four massive quarks \SeWi.  The five parameters in the
Seiberg-Witten theory are the four quark masses $\{ m_i\}$ and the
gauge coupling constant $\t_0$. The Seiberg-Witten theory
corresponding the F-theory singularity has a physical interpretation
as the world-volume theory of a D3-brane probe near the singularity
\BDS. In this correspondence, the positions of the four D-branes near
the orientifold plane are given by the squares of the four masses
and the asymptotic value $\t_0$ of the axion-dilaton field equals the
gauge coupling constant. Using this parametrization in terms of the
Seiberg-Witten curve \SeWi, the equation
\deform\ can be recast as
\eqn\deform{
y^2 = x^3 + x \a(\t_0)\ft({z}, {m_i}, \t_0) + \gt({z}, {m_i}, \t_0),}
where $\ft$ and $\gt$ are as before polynomial of degree two and
three respectively but the coefficients of the leading power of $z$
are chosen equal to one. In this picture, the orientifold plane
splits into two planes which corresponds on the probe world-volume
theory to the splitting of the origin with the $SU(2)$ classical
symmetry into a `monopole' and `dyon' point. This splitting is
non-perturbative in the string coupling and is of order
$\exp(i\pi\t_0/2)$.

The analysis near an anti-zero is the charge conjugate of the above.
For example, near $\zb=\zb_3$ we will have four anti D-branes and the
anti-orientifold plane will split into a charge-conjugate pair
corresponding to a `anti-monopole' and `anti-dyon' point on
world-volume theory of the D3-brane probe.
 
In summary, locally we have two copies of the Seiberg-Witten theory
and two copies of anti-Seiberg-Witten theory. These four can be
patched together to write the full equation that describes the most general
deformation:
\eqn\paramet{ 
y^2 = x^3 + x\, \a(\t_0) F + G,}
where
\eqn\function{\eqalign{
   F =&\, \ft(z-z_1, {m_i^1}, \t_0) \ft(z-z_2, {m_i^2}, \t_0) 
          \ft(\zb-\zb_3, {m_i^3}, \t_0)\ft(\zb-\zb_4, {m_i^4}, \t_0) \cr
   G =&\,\gt(z-z_1, {m_i^1}, \t_0) \gt(z-z_2, {m_i^2}, \t_0) 
          \gt(\zb-\zb_3, {m_i^3}, \t_0)\gt(\zb-\zb_4, {m_i^4}, \t_0) \cr}}
where $\ft$ and $\gt$ are the function appearing in equation \deform.
The overall scale of the base $\rho$ is a real parameter that is
arbitrary and not determined by this equation. Therefore, by conformal
transformation of the $\bf S^2$ base, the solution depends only on the
cross-ratio
\eqn\cross{
\omega = {(z_1-z_2)(z_3 -z_4) \over (z_1-z_3)(z_2 -z_4)}.}
The solution is thus characterized by the sixteen complex mass
parameters $\{m_i^s\}$ $(i=1,\ldots,4;s=1,\ldots,4)$, the asymptotic
value of coupling $\t_0$, the cross ratio $\omega$ and the scale $\r$
of the base. This matches with the moduli of the orientifold theory
that we have described in \S2.

Let us recapitulate our logic behind using the F-theory
description. We have established that at the special point in the
classical moduli space with $SO(8)^4$ gauge symmetry, the orientifold
theory is equivalent to the proposed F-theory. This equivalence will
hold even if we change the shape of the torus or the string coupling
in the orientifold theory as long as the tadpoles cancel locally.  The
duality thus holds on a submanifold of the moduli space with five real
dimensions where the theory has $SO(8)^4$ symmetry. In this case, the
Type-IIB supergravity equations are exactly satisfied everywhere on
the base except at the four points where the fiber degenerates giving
us a a reliable F-theory solution.  One can now turn on massless
deformations to move around the moduli space away from the submanifold
and establish the duality in a large patch of the moduli space where
the tadpoles no longer cancel locally. The branes are allowed to
wander far away from the orientifold planes. The solutions described
in this paper will be a good approximation as long as the branes are
well-separated from the anti-branes. When the branes come close to the
anti-branes, various tachyonic instabilities will kick in and the
time-independent solution that we have been discussing in terms of the
elliptic fibration will no longer be a good approximation.

Even within this approximation, the F-theory approach is more powerful
than the orientifold for studying certain aspects of the theory. For
example, as in \MoVa\ we can choose the parameters in such a way that
the solution develops two $E_8$ singularities---one from coincident
branes and the other from coincident anti-branes. We can take the
singularities to be well-separated to work within our
approximation. Thus, the theory will have $E_8\times E_8$
symmetry---something that is difficult to see in the orientifold
picture.

As we move some of the branes towards the anti-branes they will get
annihilated. Even within the orientifold, all the $D$-branes can be
annihilated along with the $\aD$-branes. We also expect that the split
$O7$ planes will annihilate the split $\aO$-planes and the final
configuration would have no charge left behind. The early stages of
this annihilation process could be analyzed systematically up to a
point using time-dependent solutions of the Type-IIB supergravity
describing motion of branes and anti-branes. However, the final stage
of this process is expected to be quite violent and will excite a lot
of closed string modes. If we maintain translational invariance along
the branes as they are annihilated, then the energy of annihilation
will get deposited onto the base. Because the base is compact, the
energy cannot be dissipated to asymptotic infinity. However, this will
generate a potential for the moduli that control the size of the base
and we expect that the theory will decompactify to restore the
supersymmetric vacuum of Type-IIB superstring.

As in the supersymmetric case, we expect that the F-theory
compactified further on a 2-torus should give a Type-IIA
compactification.  To see this explicitly, we compactify the 6th and
7th direction on circles and perform two
T-dualities\footnote{$^{\dag}$}{See for example \Dabh\ for the
T-duality transformation of $\Omega$ etc.}. This gives an orientifold
of Type-IIB with the orientifold symmetry $\{ 1, I_{6789}
\O \}\times \{1, I_{6789}\OF\hs\}$ 
which contains $D5$-branes, $O5$-planes and similarly and
$\aOf$-branes and $\aOf$-planes.  Now, we can
perform an S-duality transformation which turns $\O$ into $(-1)^{F_L}$
and we get an orbifold $\{ 1, I_{6789} \FL\}\times\{1,
I_{6789}\FR\hs\}$.  Under this duality the $D5$-branes turns into
$NS5$-branes and the orientifold planes would turn into orbifold
planes with NS charge.  Tachyons in the open string sector of the
orientifold when a $D5$ brane comes close to $\aDf$ brane correspond
in the S-dual orbifold to a {\it closed} string tachyon when
an $NS5$ to to an anti-${NS}5$ brane approaching each-other.  A
further T-duality along the 6th direction will give us a
non-supersymmetric Type-IIA compactification.

\newsec{\secfont Conclusions and Speculations}

The main conclusion of this paper is that certain nonsupersymmetric
orientifolds with tachyons can be viewed as an assembly of dynamical
charged objects with vanishing net charge.  This orientifold can be
analyzed reliably as an F-theory compactification in certain regions
of the moduli space and can be viewed as a non-supersymmetric unstable
state in the Type-IIB superstring sitting at an extremum of energy in
the configuration space.  There are a number of unstable directions at
this extremum that correspond to closed and open string tachyons.
{}From general principles, we expect this assembly to pair-annihilate
completely to relax to the vacuum containing no charges.  A detailed
study of this dynamical process is of course quite complicated but
fortunately not required for determining the final state. Thus, the
endpoint of tachyon condensation is expected to be the vacuum of
Type-IIB superstring. This is not any different from a system of equal
number of electrons and positrons---the detailed dynamics is
complicated to analyze even in QED but one can be sure that the system
would decay to the vacuum after complete pair-annihilation. One
difference for a system of 7-branes is that the charges are
non-abelian, which makes it possible to have configurations where the
net force is zero between bunches of charges and anti-charges, but
this does not alter the conclusion about the final state.

Let us compare this situation with the analysis of open-string
tachyons in the $D$-$\bar D$ system. Within the open-string theory,
there is no dynamical picture of what happens when the tachyons
condense. It is only from the viewpoint of the closed-string channel
that the D-branes are dynamical objects. This interpretation leads to
the conjecture about the endpoint of tachyon condensation which can
then be tested using open-string field theory.  Similarly, here,
within the orientifold theory, there is no obvious dynamical
interpretation of the closed string tachyons. It is only in the
F-theory dual that orientifold planes are dynamical objects and this
interpretation suggests a conjecture about the endpoint of tachyon
condensation in the orientifold theory. The analysis of the F-theory
dual is based on the supergravity equations which are the low energy
limit of the full closed string field theory. To study the dynamical
process of tachyon condensation would require a time-dependent
F-theory like solution of the Type-IIB supergravity equations and
moreover in the regions of interest the $\apm$ corrections will become
important. It would be interesting to see if closed string field
theory or a variant of the methods utilized in \APS\ for studying
twisted sector tachyons can shed more light on this process.

We end with some speculative remarks about the A-orientifold mentioned
in \S2 which was part of the motivation for the present work.  The
action of our orientifold was cleverly chosen so that the $O7$-planes
were separated from the $\aO$-planes. We could reliably study the
F-theory dual when this separation was large and establish that the
orientifold plane would split into two dynamical 7-branes using local
analysis. In this regime there are no tachyons in the spectrum. By
contrast, for the A-orientifold with orientifold symmetry $\{ 1,
\OL\II\}\times\{ 1, \OR\II\}$, the $O7$ planes are coincident with the
$\aO$-planes and the closed-string tachyons are always present (Fig 2).
\fig{The crosses indicating the orientifold planes
and the circles indicating the anti-orientifold planes now coincide.
The fundamental region of the orientifold is half of the original
torus that is inside dashed lines.}{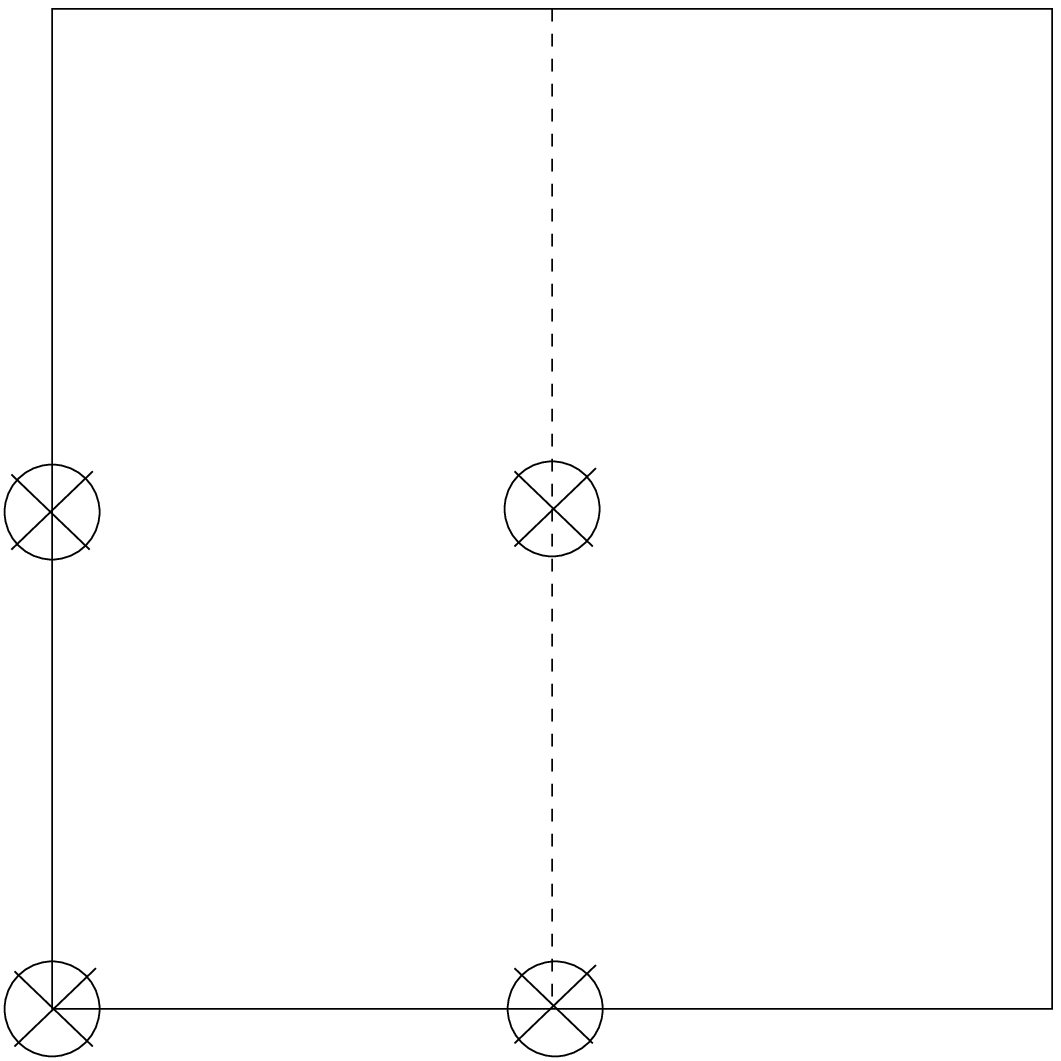} {1.0truein}
\noindent There is no
systematic way to analyze this situation in F-theory without dealing
with tachyons but it is natural to speculate that even in this case
the $O7$ and the $\aO$ planes would split nonperturbatively and would
have the interpretation of dynamical objects.  It is then reasonable
to assume that this system will also dynamically pair-annihilate
completely into the Type-II vacuum. Indeed, a comparison of Figs 1 and
2 suggests that we may be able to view the A-orientifold as a double
cover of our F-theory with half-branes stuck at the fixed points.
Note that the T-dual of the A-orientifold has been conjectured to be
S-dual to the bosonic string. There are also intriguing relations
between the bosonic string and the superstring
\CENT\ that underlie the `bosonic map'\LSW. If the A-orientifold is
indeed dual to the bosonic string, then it leads to a speculation
that the 26-dimensional bosonic string decays to the Type-II
superstring. At this stage, however, we do not have more definitive
evidence to offer.

There are a number of other recent conjectures about the fate of
related non-supersymmetric theories such as the 0B theory, its
orientifolds, and heterotic duals \refs{\CoGu,
\GuSt, \Suya}. These are based on the interpretation of the
0B theory in terms of the Melvin solution of the Type-IIB string.
Thus, also from this very different point of view, the 0B theory and
possibly its orientifolds appear as excited configurations of the
Type-IIB string which are then expected to decay to the supersymmetric
vacuum. It would be interesting to see how these ideas connect with
the those put forward in this paper.

\leftline{ \secfont Acknowledgments}
I would like to thank Shiraz Minwalla and Sandip Trivedi for
stimulating discussions.
\bigskip
\vfill
\eject
\listrefs
\end